\documentclass{PoS}

\usepackage{bm}

\title{Semileptonic form factor ratio $B_s\to D_s/B\to D$ and its application to $BR(B^0_s\to \mu^+\mu^-)$}  

\ShortTitle{Semileptonic form factor ratio $B_s\to D_s/B\to D$}

\author{
\speaker{Daping Du}$^{,a,b,c}$, Carleton DeTar$^d$, Andreas Kronfeld$^b$, Jack Laiho$^e$, \quad\quad\quad\quad\quad
    Yannick Meurice$^a$, and Si-wei Qiu$^d$ \\ 
    $^a$Department of Physics and Astronomy, University of Iowa, Iowa City, IA 52240, USA\\
    $^b$Fermi National Accelerator Laboratory, Batavia, IL 60510, USA \\
    $^c$Physics Department, University of Illinois, Urbana, IL 61801, USA \\
    $^d$Department of Physics and Astronomy, University of Utah, Salt Lake City, UT 84112, USA\\
    $^e$SUPA, Department of Physics and Astronomy, University of Glasgow, Glasgow, Scotland, UK \\
    E-mail: \email{ddu@illinois.edu}}
\author{Fermilab Lattice and MILC Collaborations}

\abstract{We present a (2+1)-flavor lattice QCD calculation of the form factor ratio between the 
semileptonic decays $\bar{B}^0_s \to D^+_sl^-\bar{\nu} $ and $\bar{B}^0 \to D^+l^-\bar{\nu} $. This ratio is 
an important theoretical input to the hadronic determination of the $B$ meson fragmentation fraction ratio 
$f_s/f_d$ which enters in the measurement of $\mathrm{BR}(B^0_s\to \mu^+\mu^-)$. Small lattice spacings and 
high statistics enable us to simulate the decays with a dynamic final $D$ meson of small momentum and 
reliably extract the hadronic matrix elements at nonzero recoil. We report our preliminary result for the 
form factor ratio at the corresponding momentum transfer of the two decays 
$f_0^{(s)}(M^2_\pi)/f_0^{(d)}(M^2_K)$.} 

\FullConference{XXIXth International Symposium on Lattice Field Theory\\
                July 10-16, 2011\\
                Squaw Valley, Lake Tahoe, California}

\begin{document}

\newcommand{\half}{\frac{1}{2}}
\newcommand{\bea}{\begin{eqnarray}}
\newcommand{\eea}{\end{eqnarray}}
\newcommand{\bdm}{\begin{displaymath}}
\newcommand{\edm}{\end{displaymath}}
\newcommand{\<}{\langle}
\renewcommand{\>}{\rangle}
\newcommand{\Tr}{\mbox{Tr}}
\newcommand{\be}{\begin{equation}}
\newcommand{\ee}{\end{equation}}
\newcommand{\ben}{\begin{eqnarray}}
\newcommand{\een}{\end{eqnarray}}

\vspace*{-0.5in}
\section{Introduction}                      
The rare decay $B^0_s\to \mu^+\mu^-$ is a process that is potentially sensitive to  physics beyond the 
standard model (SM). In the SM, the decay can go only through penguin or box topologies at the loop level. 
Thus, a small branching fraction has been predicted, with the aid of lattice QCD, to be $3.2(2)\times 10^{-9}$ 
\cite{Gamiz:2009ku,Buras:2003td}. Recently, LHC$b$ \cite{2011arXiv1103.2465T} and CDF \cite{Aaltonen:2011fi} reported 
bounds on the branching fraction, to be followed by upcoming results from CMS.
It is likely that a $5\sigma$ measurement will be made, even at the SM branching ratio, in the near future.

At LHC$b$, the extraction of the branching fraction relies on the normalization channels $B_u^+\to J/\psi 
K^+$, $B^0_d\to K^+\pi^-$ and $B^0_s\to J/\psi \phi$ \cite{:2009ny}, through the following relation
\be
    \mathrm{BR}(B_s^0 \to \mu^+\mu^- ) = \mathrm{BR}(B_q\to X)\frac{f_q}{f_s} 
        \frac{\epsilon_X}{\epsilon_{\mu\mu}} \frac{N_{\mu\mu}}{N_X},
\ee
where $\epsilon$ and $N$ are the detector efficiencies and number of events.
The fragmentation fractions, $f_q$ ($q=u,d,s$ or $\Lambda$), denote the probability of a $b$ quark 
hadronizing into a $B_q$ meson or a $b$-flavored (e.g., $\Lambda_b$) baryon.
The fragmentation fraction ratio $f_s/f_d$ is crucial
in the extraction of $\mbox{BR}(B_s^0 \to \mu^+\mu^- )$.
Currently, the uncertainty in $f_s/f_d$ is the major source of uncertainty. Traditionally, $f_s/f_d$ was 
measured using the ratio of the corresponding semileptonic decays. Fleischer, Serra and Tuning proposed 
\cite{Fleischer:2010ay} that the ratio can also be measured using the non-leptonic decays $\bar{B}^0_s\to 
D_s^+\pi^-$ and $\bar{B}^0_d \to D^+K^-$, which has the advantages of a cleaner background, similar 
reconstruction of final states, {\it etc}. The approach is based on factorization of the nonleptonic 
amplitudes into $f_\pi$ or $f_K$ and corresponding semileptonic form factors. The ratio $f_s/f_d$ is related 
to $\mbox{BR}(\bar{B}_s\to D\pi)/\mbox{BR}(\bar{B}\to DK)$ in a way similar to Eq.~(1.1).  With the 
efficiencies and event counts combined with the factorization approximation, we have
\be
    \frac{f_s}{f_d} =  \frac{|V_{us}|^{2}f_K^2}{|V_{ud}|^{2}f_\pi^2} \times	
        \frac{\tau_{B^0}}{\tau_{B_s^0}} \times 
        \left [ \frac{\epsilon_{DK}}{\epsilon_{D_s\pi}}\frac{N_{D_s\pi}}{N_{DK}}\right ]\times
        \frac{1}{\mathcal{N}_a\mathcal{N}_F}, 
\ee
where $\tau$ is the lifetime and $\mathcal{N}_a\approx 1$ with corrections of a few percent due to 
nonfactorizable effects \cite{Fleischer:2010ay}. The semileptonic form factor ratio $\mathcal{N}_F 
=[f_0^{(s)}(M^2_\pi)/f_0^{(d)}(M^2_K) ]^2$ is currently the decisive contributor to the theoretical error. 
The value currently used at LHC$b$ is an estimate from QCD sum rules, $\mathcal{N}_F = 1.24(8)$ 
\cite{Blasi:1993fi, Fleischer:2010ay}. However, this theoretical input and the size of its error need to be 
validated by a nonperturbative method such as lattice QCD. This paper is devoted to such a calculation.

The matrix elements of the $B_{}\to D_{}$ semileptonic decay (and similarly for $B_s\to D_s$) can be written 
as
\be
    \langle D(p') | \mathcal{V}^\mu | B(p) \rangle =  
        f_+ (q^2) \left [ (p+p')^\mu - \frac{M_B^2-M_D^2}{q^2}  q^\mu \right ] +  
        f_0(q^2) \; \frac{M_B^2-M_D^2}{q^2} q^\mu.
    \label{eq:f0f+}
\ee
However, for heavy quarks it is convenient to use the variables $h_\pm$, defined by
\be
    \frac{\langle D(p') | \mathcal{V}^\mu | B(p) \rangle }{ \sqrt{M_B M_D}} = 
        h_+(w) \; (v + v')^\mu  +  h_-(w)\; ( v-v')^\mu , 
    \label{eq:h+h-}
\ee
where $v = p/M$ and the recoil variable is $w = v \cdot v'$ . 
We will use the form factors $h_\pm$ in our entire analysis and convert them in the end to $f_+, f_0$ using 
Eqs.~(\ref{eq:f0f+}) and~(\ref{eq:h+h-}). 

In these proceedings,  we report a preliminary result of the form factor ratio 
$f_0^{(s)}(M^2_\pi)/f_0^{(d)}(M^2_K)$ by analyzing the semileptonic decays $\bar{B}^0_s \to 
D^+_sl^-\bar{\nu} $ and $\bar{B}^0 \to D^+l^-\bar{\nu} $ on the lattice. \pagebreak  We use an identical subset of the 
MILC gauge configurations for both of the $B_s\to D_s$ and $B\to D$ processes. To reduce the statistical 
errors effectively, we construct a set of ratios at small recoil, from which we extract the lattice form 
factors $h_\pm$. The extrapolation to physical light quark masses and to the continuum is
performed using root staggered chiral perturbation theory (rS$\chi$PT). 
The results are extrapolated to maximum recoil by employing a model-independent parametrization.
In Sec.\ref{sec:4}, we report our lattice result.

\vspace*{-0.1in}
\section{Numerical details}  
\vspace*{-0.1in}
\subsection{Data setup}
Our calculation uses four ensembles of the MILC's (2+1)-flavor gauge configurations \cite{Bazavov:2009bb}, two 
at each of the lattice spacings $a \approx 0.12$~fm and $\approx 0.09$~fm. The ensembles as well as the 
parameters used are summarized in Table~\ref{tab:ensembles}.  The strange and light sea quarks were simulated using the 
asqtad-improved staggered action \cite{Lepage:1998vj}. The action is also used in our strange and light 
valence quarks. The heavy quarks (charm and bottom) are simulated using the Sheikholeslami-Wohlert (SW) 
clover action  with the Fermilab interpretation \cite{ElKhadra:1996mp}.  For the $B\to D$ 
decay, the spectator light quark is degenerate with the light sea quark (full QCD). While for the $B_s\to 
D_s$ decay, the strange quark is set close to its physical value. The charm and bottom quarks in our 
calculation are tuned to their physical values up to a tuning uncertainty. The corresponding bare hopping 
parameter $\kappa_{b(c)}$, as well as the coefficient for the clover term $c_{\rm SW}$ are given explicitly in 
Table~\ref{tab:ensembles}.
\begin{table}[b]
\vspace{-0.1in}
\begin{tabular}{cccccccc}
\hline
\hline
$a$~(fm) & 
$am_l/am_s$ & 
$ N_{\mbox{\scriptsize confs}}$ & 
$c_{\rm SW}$  &
$\kappa_c$  &
$\kappa_b$  & 
$am_x(B\to D)$ & 
$am_x(B_s\to D_s)$\cr
\hline 
$\approx$0.12&0.020/0.050 & 2052 & 1.525 &0.1259& 0.0918  & 0.020 & 0.0349\cr 
$\approx$0.12&   0.007/0.050 & 2110 & 1.530 &0.1254& 0.0901 & 0.007 & 0.0349\cr 
$\approx$0.09&0.0124/0.031 & 1996 & 1.473 &0.1277& 0.0982& 0.0124 & 0.0261\cr 
$\approx$0.09&0.0062/0.031 & 1931 & 1.476 &0.1276& 0.0979 & 0.0062 & 0.0261\cr
\hline
\hline
\end{tabular}
\caption{\label{tab:ensembles} MILC ensembles of configurations used in this analysis. }
\end{table}

\vspace{-0.1in}
\subsection{Lattice extraction} 

In this work, we are interested only in the vector current operator. On the lattice we define
$V^\mu = \sqrt{Z^{cc}_{V^4} Z^{bb}_{V^4}} \; \overline{\Psi}_c i\gamma^\mu \Psi_b $,
where $Z^{hh}$ are normalization factors. The vector current in the continuum is 
$\mathcal{V}^\mu = \rho_{V^\mu} V^\mu$, where $\rho_{V^\mu}^2 = 
Z^{bc}_{V^\mu}Z^{cb}_{V^\mu}/Z^{bb}_{V^4}Z^{cc}_{V^4}$. The factor $\rho_{V^\mu}$ can be calculated 
perturbatively and has been found to be very close to one \cite{Kronfeld:2000ck,Harada:2001fj}. 
We expect the $\rho_V$s to largely cancel in the ratio of the form 
factors. Hence this correction is negligible compared with other systematic errors and we take $\rho_V=1$ 
for this analysis.

We employ the following three-point functions in our analysis,
\ben
C^{DV^\mu B}_{3pt}(0,t,T;{\bf p}_D) &=& \sum_{{\bf x},{\bf y}} \; \langle 0| \mathcal{O}_D(0, {\bf 0})  \overline{\Psi}_c i\gamma^\mu \Psi_b(t, {\bf y}) \mathcal{O}^+_B(0, {\bf x}) | 0 \rangle \; e^{i {\bf p}_D\cdot{\bf y}},  \\
C^{DV^\mu D}_{3pt}(0,t,T;{\bf p}_D) &=& \sum_{{\bf x},{\bf y}} \; \langle 0| \mathcal{O}_D(0, {\bf 0})  \overline{\Psi}_c i\gamma^\mu \Psi_c(t, {\bf y}) \mathcal{O}^+_D(0, {\bf x}) | 0 \rangle \; e^{i {\bf p}_D\cdot{\bf y}}, \\
C^{BV^4 B}_{3pt}(0,t,T;{\bf 0}) &=& \sum_{{\bf x},{\bf y}} \; \langle 0| \mathcal{O}_B(0, {\bf 0})  \overline{\Psi}_b i\gamma^4 \Psi_b(t, {\bf y}) \mathcal{O}^+_B(0, {\bf x}) | 0 \rangle .
\een
\vspace*{-0.2in}

\noindent
The $B$ meson is at rest. 
To obtain the dependence of the form factors at small recoil $w$, we simulate the final \pagebreak
state $D$ meson at a few small momenta, {\it i.e.}, ${\bf p} = 2\pi(1,0,0)/L$ ,  
$2\pi(1,1,0)/L$,  $2\pi(1,1,1)/L$ and $ 2\pi(2,0,0)/L$. 
The correlation functions for $D\to D$ and $B\to B$ serve as normalization. 
The $D\to D$ correlation function with a non-zero final state momentum is used to extract the recoil $w$ to 
alleviate the need of renormalizing the four velocity.

From these correlation functions we construct three different ratios, fits to which include contributions 
from the lowest-order excited states. Explicitly, 
\ben
\frac{C^{DV^iD}_{3pt}(0,t,T;{\bf p})}{C^{DV^4D}_{3pt}(0,t,T;{\bf p})} &=& d^i\;(1 + \mathcal{D}_{02}\;e^{- \Delta E({\bf 0})(T-t)} + \mathcal{D}_{20}\;e^{- \Delta E({\bf p}) t}) \label{eq:di_fit},\\
\frac{C^{DV^iB}_{3pt}(0,t,T;{\bf p})}{C^{DV^4B}_{3pt}(0,t,T;{\bf p})} &=& b^i\;(1 + \mathcal{B}_{02}\;e^{- \Delta m(T-t)} + \mathcal{B}_{20}\;e^{- \Delta E({\bf p}) t}) \label{eq:bi_fit},\\
 \frac{C^{DV^iB}_{3pt}(0,t,T;{\bf p})}{C^{DV^4B}_{3pt}(0,t,T;{\bf 0})}&\times& \left [\frac{Z_0({\bf 0})}{Z_0({\bf p})}\sqrt{\frac{E_0({\bf p})}{E_0({\bf 0})}}  e^{(E_0({\bf p})-E_0({\bf 0})) t} \right ]  \nonumber \\
&=& a^i\;(1 + \mathcal{A}_{02}\;e^{- \Delta m(T-t)} + \mathcal{A}_{20}\;e^{- \Delta E({\bf p}) t} + \mathcal{A'}_{20}\;e^{- \Delta E_2({\bf 0}) t}) e^{\delta t} .\label{eq:ai_fit}
\een
The factor in the square brackets of Eq.~(\ref{eq:ai_fit}) cancels the time dependence of 
the ratio, stemming from the fact that the numerator and denominator have final state $D$ mesons with 
different momenta. $\Delta E$ and $\Delta m$ denote the lowest splittings and $\delta$ is a parameter that 
accounts for the imprecise $E({\bf p})-E({\bf 0})$ in the bracket in Eq. (\ref{eq:ai_fit}). 
In the fits the lowest-lying energy splittings $\Delta E$, $\Delta m$ are treated as fit parameters. 
The splittings can be extracted from the two-point functions. So, we employ a multi-channel fitting procedure, 
combining the two-point functions and the ratios of the three-point functions. We find that such a treatment 
results in more robust fits and more precise splittings. From $d_i, b_i$ and $a_i$ we can easily recover the 
form factors $h_\pm$ at small recoil $w$,
\ben
w &=& \frac{1+{\bf d}\cdot{\bf d}}{1-{\bf d}\cdot{\bf d}}, \\
h_+(w) &=&   h_+(1)(a_i/b_i-{\bf a}\cdot{\bf d}),\\
h_-(w) &=&   h_+(1)(a_i/b_i-a_i/d_i).
\een

\vspace*{-0.3in}
\section{Results}     
\vspace*{-0.1in}

The extrapolation of our lattice results to the physical quark masses and the continuum is guided by 
rS$\chi$PT \cite{Aubin:2003mg,Laiho:2005ue}. However, in the case of $h_+$ the light quark mass dependence 
is accompanied by a small recoil $w$ dependence. Such a dependence was included in the continuum chiral 
perturbation theory in Ref.~\cite{Chow:1993hr} and was extended to the NLO rS$\chi$PT in 
Ref.~\cite{Laiho:2011}. For $h_-$, the NLO correction is simply a constant which is inversely proportional 
to the charm quark mass. We follow the same setup, adding NNLO analytic terms and including $a^2$ 
dependence. The remaining recoil dependence of the form factors is fitted to a simple quadratic expansion at 
zero recoil.

\begin{figure}[b]
    \vspace{-0.4em}
\begin{minipage}[b]{0.5\linewidth} 
\centering
\includegraphics[width=\textwidth, trim=10mm 0 0 0]{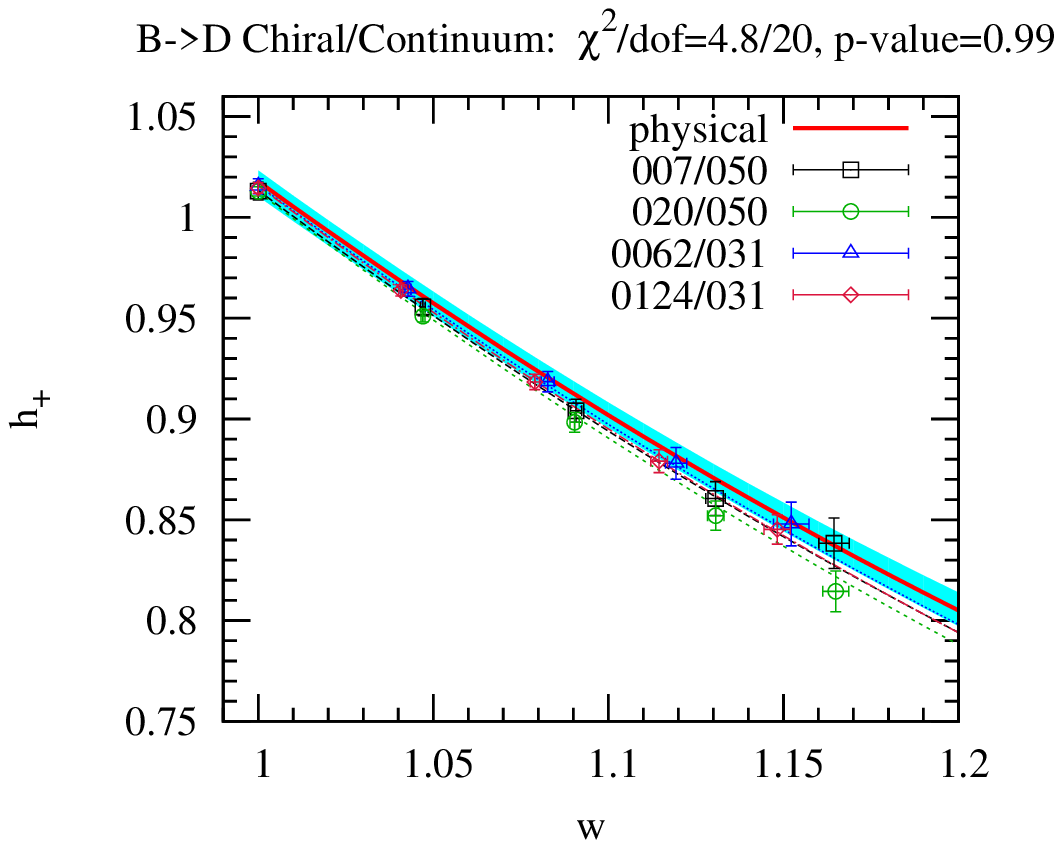}
\end{minipage}
\hspace{0.5cm}
\begin{minipage}[b]{0.5\linewidth}
\centering
\includegraphics[width=\textwidth, trim = 10mm 0 0 0]{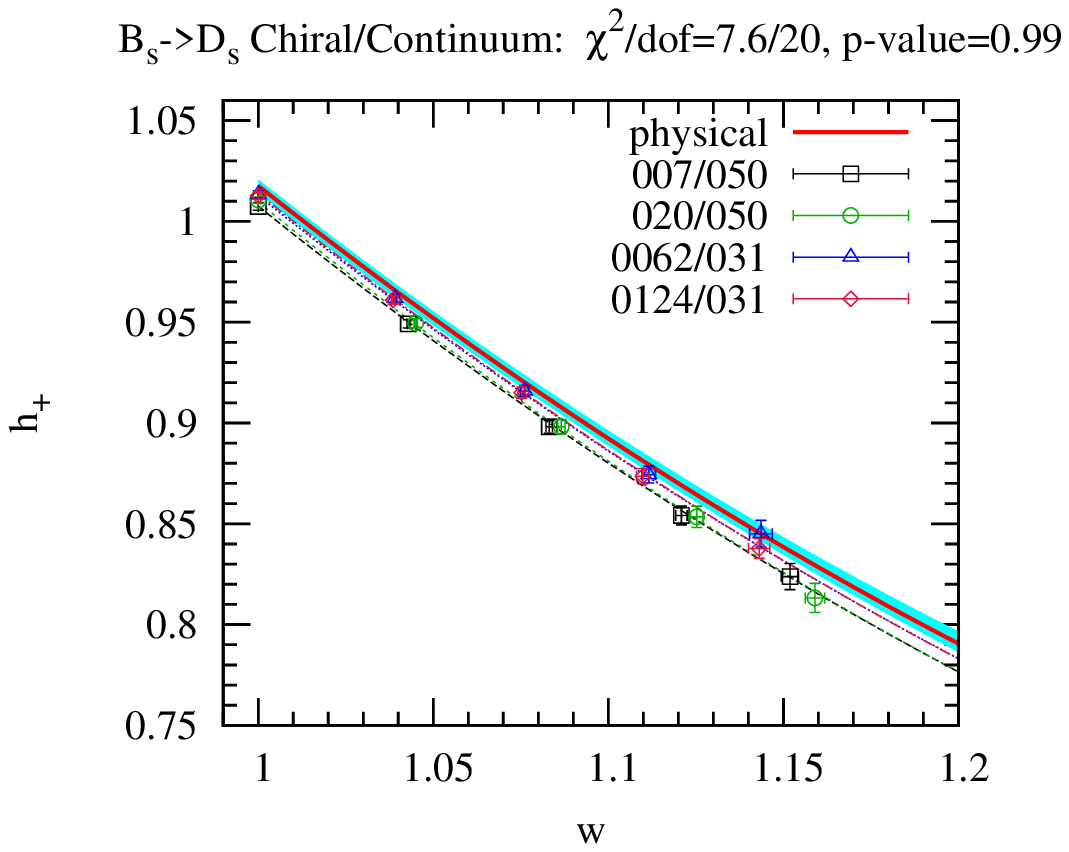}
\end{minipage}
\begin{minipage}[b]{0.5\linewidth} 
\centering
\includegraphics[width=\textwidth, trim=10mm 0 0 0]{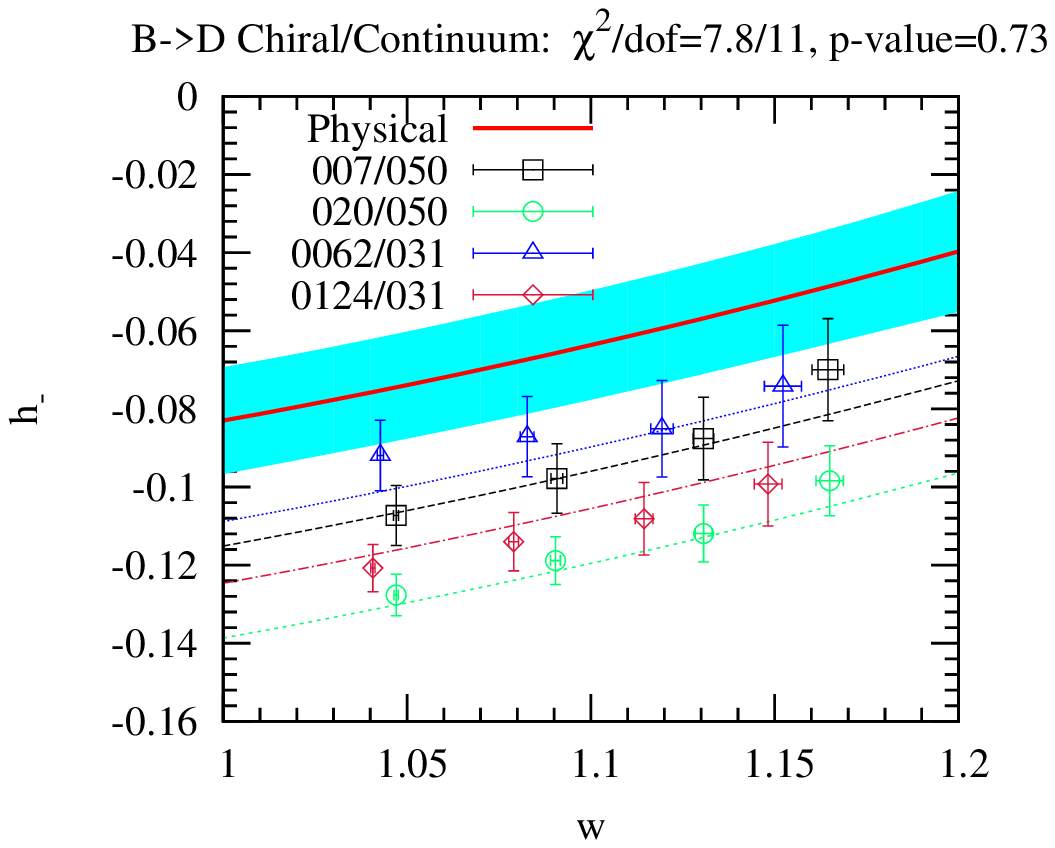}
\end{minipage}
\hspace{0.5cm}
\begin{minipage}[b]{0.5\linewidth}
\centering
\includegraphics[width=\textwidth, trim = 10mm 0 0 0]{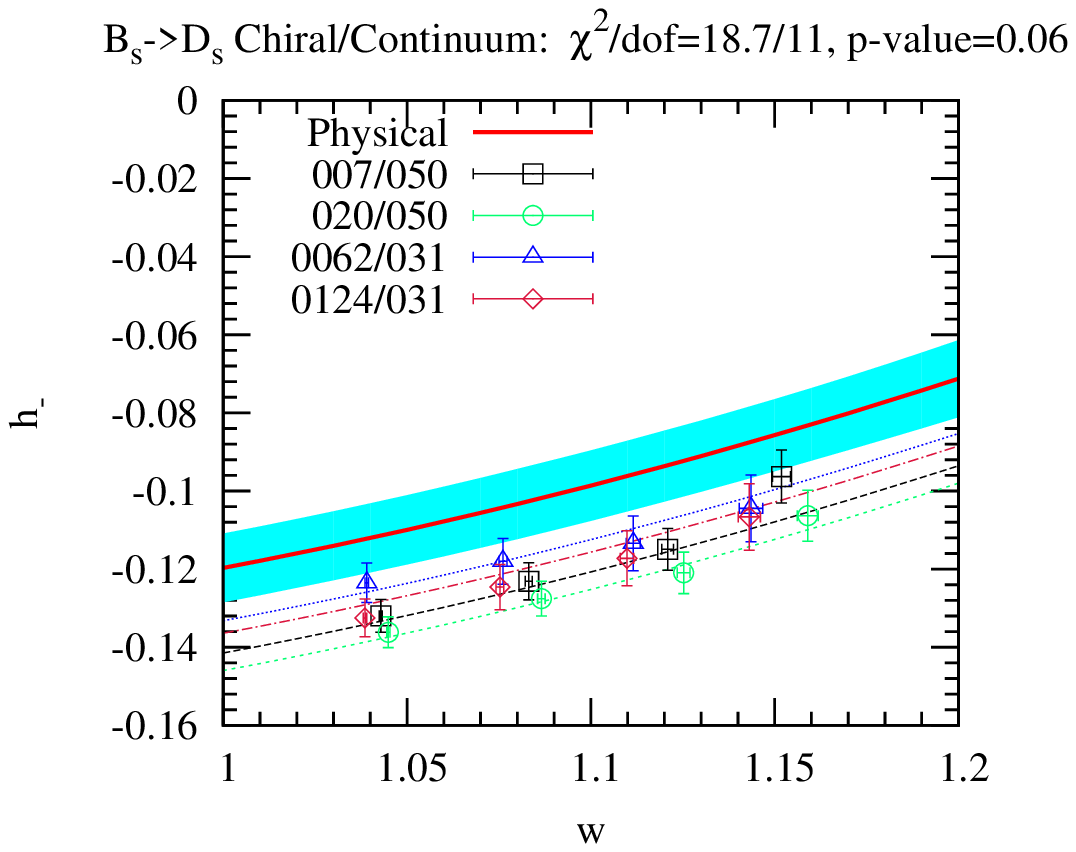}
\end{minipage}
\caption{Chiral/continuum extrapolation of $h_\pm(w)$ for the $B\to D$ (left) and $B_s \to D_s$ decays 
(right). \label{fig:chiral}} 
\end{figure}

The results of the chiral/continuum extrapolation are shown in Fig.~\ref{fig:chiral}.
The form factor $h_+$ for both of the $B\to D$ and $B_s\to D_s$ decays shows a small dependence on the light 
quark masses and lattice spacings. The extrapolated physical values are very close to the lattice data 
points. This suggests that $h_+$ is insensitive to the light degrees of freedom. However, sizable light 
quark mass and lattice spacing dependence appears in the case of $h_-$, as indicated by the variation due to 
the sea quark masses and the differences between $h_-^{B\to D}$ and $h_-^{B_s\to D_s}$ (spectator mass). 
Note that the difference between $h_+^{B\to D}$ and $h_+^{B_s\to D_s}$ is minor. Considering the subleading 
role that $h_-$ plays in contributing to $f_+, f_0$, we expect the U-spin symmetry breaking effect to be 
\pagebreak
smaller than what was expected in \cite{Fleischer:2010ay,Blasi:1993fi}. Such an observation is bolstered by 
the recent lattice calculations on $f_+, f_0$ of the $D_{(s)}\to \pi (K)$ decays \cite{Koponen:2011ev}.

\begin{figure}[h]
\begin{minipage}[b]{0.5\linewidth} 
\centering
\includegraphics[width=\textwidth, trim=10mm 0 0 0]{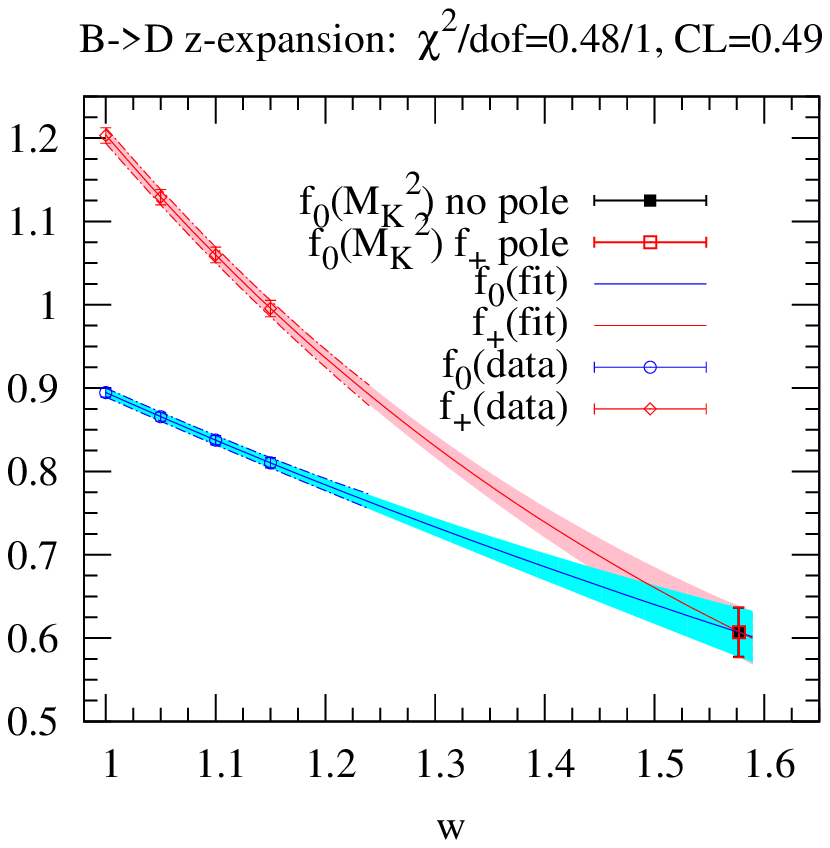}
\end{minipage}
\hspace{0.5cm}
\begin{minipage}[b]{0.5\linewidth}
\centering
\includegraphics[width=\textwidth, trim = 10mm 0 0 0]{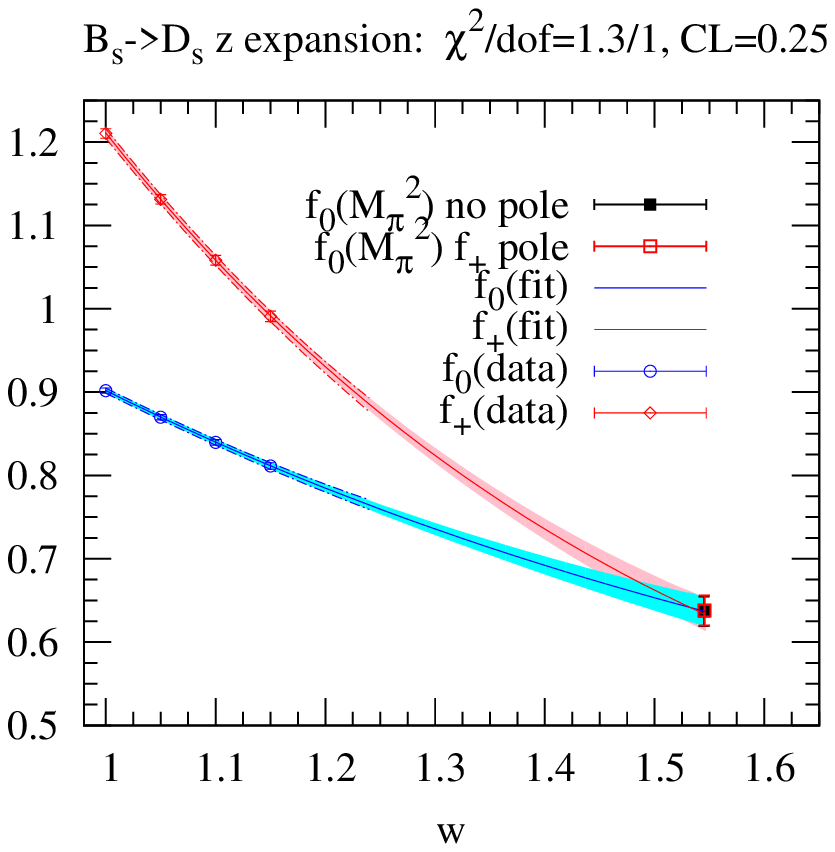}
\end{minipage}
\caption{ The z-expansion of form factors $f_0, f_+$. The points that we include in the z-expansion fits are shown explicitly. The dashed curves indicate the result of chiral/continuum extrapolation. \label{fig:zfit} }
\end{figure}

With the physical values of $h_\pm$, we can easily calculate $f_0, f_+$ using the physical masses of the $D$ and $B$ mesons. However, to evaluate the form factors at a small momentum transfer ($q^2 = M_\pi^2, M_K^2$), we need to extrapolate the results near maximum recoil. We use the model-independent $z$-parametrization \cite{Boyd:1994tt} with the constraint $f_0(0) = f_+(0)$. We take four synthetic points in the recoil range where we have lattice data points. We take the values of $f_+, f_0$ by evaluating our chiral/continuum extrapolation result at these four recoil points and perform the z-parametrization. The result is shown in Figure \ref{fig:zfit}. We study the effect of a pole at a vector $B_c$ meson in the Blascke factor of the $z$-expansion of $f_+$. We find that the shapes of the form factors are only weakly affected by the inclusion of such a pole. 

By expanding the form factors at the respective momentum transfers, we finally arrive at 
\be
f_0^{(s)}(M^2_\pi)/f_0^{(d)}(M^2_K)=1.051(40)(22).
\ee
The first error is from statistics. The second error is the systematical error due only to the uncertainty on $g_{DD^*\pi}$ and to the variation of fits in the $z$-parametrization. We are in the process of building a full systematic error budget. 
\pagebreak

\vspace*{-0.1in}
\section{Conclusions}              
\label{sec:4}
\vspace*{-0.1in}
In summary, we present a (2+1)-flavor lattice QCD calculation of the form factor ratio\\ $f_0^{(s)}(M^2_\pi)/f_0^{(d)}(M^2_K)$, which is a major theoretical input for the extraction of the fragmentation fraction ratio $f_s/f_d$. The essential part of our calculation is to extract the $B\to D$ and $B_s\to D_s$ semileptonic form factors at non-zero recoil. We reduce the systematic uncertainty by fitting the lowest-order excited states, and we employ a simultaneous multi-channel fit procedure to address correlations and reduce the statistical uncertainty. Our chiral/continuum results show that the corrections to the finite lattice spacings and finite light quark masses are small. Our preliminary result is $f_0^{(s)}(M^2_\pi)/f_0^{(d)}(M^2_K)=1.051(40)(22)$, with a partial systematic error budget. As a consequence, we obtain $\mathcal{N}_F = 1.105(80)(44)$ which implies a smaller U-spin breaking effect than that suggested in \cite{Blasi:1993fi}, $\mathcal{N}_F = 1.24(8)$. A more comprehensive analysis with a detailed error budget is still in progress and will be reported in a forthcoming paper. 

\medskip

D.D. thanks Aida El-Khadra for several helpful discussions on the perturbation theory. D.D. 
has also received considerable help from Chris Bouchard, Elizabeth Freeland, James Simone, Jon Bailey, 
Elvira Gamiz and Ran Zhou, who shared their numerical techniques or codes. Computations for this work were 
carried out with resources provided by the USQCD Collaboration, the Argonne Leadership Computing Facility, 
the National Energy Research Scientific Computing Center, and the Los Alamos National Laboratory, which are 
funded by the Office of Science of the U.S. Department of Energy; and with resources provided by the 
National Center for Supercomputer Applications, the National Institute for Computational Science, the 
Pittsburgh Supercomputer Center, the San Diego Supercomputer Center,and the Texas Advanced Computing Center, 
which are funded through the National Science Foundation Teragrid/XSEDE Program. D.D. was supported in part 
by the URA Visiting Scholars' program at Fermilab. This work was supported in part by the U.S. Department of 
Energy under Grants No.~DE-FG02-91ER40664 (Y.M., D.D.), DE-FE06-ER41446 (C.D.), No.~DE-FG02-91ER40677 (D.D) 
and in part by the U.S. National Science Foundation under Grants PHY0757333 (C.D.) and PHY0903571 (S.-W.Q.). 
J.L. is supported by the STFC and by the Scottish Universities Physics Alliance. Fermilab is operated by 
Fermi Research Alliance, LLC, under Contract No. DE-AC02-07CH11359 with the United States Department of 
Energy.

\bibliographystyle{JHEP2a}
\bibliography{Bs_project}

\end{document}